\documentclass[aps,prb,twocolumn,groupedaddress,showpacs]{revtex4}
\usepackage{epsfig}
\usepackage[dvipsnames,usenames]{color}
\usepackage{color}
\usepackage{amsmath}
\usepackage{amssymb}

\emergencystretch=\maxdimen
\begin{document}
\title{Interaction-driven Band-insulator--to--Metal transition in bilayer ionic Hubbard model}
\author{M. Jiang$^1$ and T.C.S. Schulthess$^{1,2}$}

\affiliation{$^1$Institute for Theoretical Physics, ETH Zurich, Switzerland}
\affiliation{$^2$Swiss National Supercomputing Center, ETH Zurich, 6900 Lugano, Switzerland}

\begin{abstract}
The interaction-driven insulator-to-metal transition has been reported in the ionic Hubbard model (IHM) for moderate interaction $U$, while its metallic phase only occupies a narrow region in the phase diagram. To explore the enlargement of the metallic regime, we extend the ionic Hubbard model to two coupled layers and study the interplay of interlayer hybridization $V$ and two types of intralayer staggered potentials $\Delta$: one with the same (in-phase) and the other with a $\pi$-phase shift (anti-phase) potential between layers. Our determinant Quantum Monte Carlo (DQMC) simulations at lowest accessible temperatures demonstrate that the interaction-driven metallic phase between Mott and band insulators expands in the $\Delta-V$ phase diagram of bilayer IHM only for in-phase ionic potentials; while anti-phase potential always induces an insulator with charge density order. This implies possible further extension of the ionic Hubbard model from the bilayer case here to realistic three-dimensional model. 
\end{abstract}
\pacs{71.10.Fd, 71.30.+h, 02.70.Uu}
\maketitle

\section{Introduction}
The ionic Hubbard model (IHM) with alternating site energies has been investigated as an effective Hamiltonian due to its relevance to various phenomena in correlated electronic systems such as the evolution of electronic structure in SrRu$_{1-x}$Ti$_{x}$O$_{3}$\cite{realsystem0}, enhanced response of quasi-one-dimensional ferroelectric perovskites\cite{realsystem1}, metal to charge-transfer insulator transitions in A$_{x}$CoO$_{2}$ (A = Na, Rb, K)\cite{triangular}, and unconventional spin-singlet superconductivity in layered nitrides $\beta$-MNCl (M = Hf, Zr) \cite{honey}. Theoretically,
the electron correlation is normally believed to drive the phase transitions from metals to magnetically ordered states and Mott insulating behavior in various tight-binding Hamiltonians; while the ionic potential results in the band insulating phases on half-filled bipartite lattices. Interestingly, some theoretical studies have revealed a correlation induced intermediate state in the transition from a band insulator (BI) to a Mott insulator (MI) with the aid of ionic potentials for moderate values of Hubbard interaction\cite{RTS2007,QMC,bondorder,bondorder2015,ionic1,ionic2,ionic3,ionic4,ionic5}.
Such intermediate phases between BI and MI have been extensively investigated in other dimensions as well. For example, in one dimension, an intermediate bond-ordered phase has been reported while the metallic phase shrinks to only one point\cite{1Dionic1,1Dionic2}, which can be recovered by introducing additional intra-sublattice hopping\cite{1Dionic3}. In high dimensions $D>2$, various studies employing single-site or cluster dynamical mean field theory (DMFT)\cite{DMFTreview,DCAreview} provided fruitful insights on the competition between BI and MI\cite{bondorder,ionic1,ionic2,ionic3,ionic4}.
Intriguingly, the IHM in honeycomb lattice has also been realized using the interacting two-component gas of fermionic atoms loaded into an optical lattice recently\cite{coldatom}, which provides a cleaner platform compared with real materials for addressing open questions concerning the debated nature of the intermediate regime between the two insulating phases held in IHM\cite{RTS2007,QMC,bondorder,bondorder2015,ionic1,ionic2,ionic3,ionic4,ionic5}.

Although the intermediate states between MI and BI due to the ionic potentials poses fundamental interest in the correlated electronic systems, it only occupies a narrow region in the phase diagram, which casts additional difficulty on the investigation of its properties. Another natural question concerns the robustness of this intermediate phase against external perturbations, which in turn will shed light on the understanding of the phase itself. Motivated by these considerations, we explore the possibility of enlarging the intermediate regime in the parameter space by extending IHM to the bilayer case, in which the additional energy scale -- interlayer hybridization $V$ -- provides another degree of freedom as a tuning parameter. Treating $V$ as an external perturbation, this bilayer model also allows us to study the robustness of the intermediate phase as well, although some of its properties will be intertwined with those in the conventional bilayer Hubbard model. Another motivation for this extension to bilayer IHM originates from the fact that diverse real materials have three-dimensional lattice structures\cite{realsystem0,realsystem1,triangular,honey}, despite that 2D lattice models are conventionally assumed to be able to capture their essential physics. Since the intermediate phase occurs for moderate Hubbard interaction, which usually imposes strong finite-size effects in numerical simulations, we have to adopt a large enough two-dimensional lattice but sacrifice the length of the third dimension due to the limitation of the computational resource. Therefore, this bilayer ionic Hubbard model can be regarded as the simplest case of more realistic 3D ionic model or two-orbitial/band systems.

The bilayer IHM introduces additional complication that each layer can experience the same or different ionic potentials. This paper addresses two characteristic cases, namely two layers feel (a) the same potential (``in-phase'') and (b) the same potential amplitude $\Delta$ but with a $\pi$-phase shift (``anti-phase'').
By addressing the fate of the intermediate phase in the presence of interlayer hybridization $V$ for these two cases, we provide evidence that the intermediate metallic regime can extend to considerable regions in the $\Delta-V$ phase diagram at the lowest accessible temperature of our simulations only for in-phase ionic potentials, which implies possible further extension to realistic three-dimensional model. 
However, anti-phase ionic potential always induces an insulator with charge density order due to the coupling of potential peak and valley between layers.

\section{Model and methodology}

The bilayer ionic Hubbard model reads as
\begin{equation}
\begin{split}
\hat{H} = &-t \sum\limits_{\langle \mathbf{r}\mathbf{r'} \rangle m \sigma}
(c^{\dagger}_{\mathbf{r}m\sigma}c_{\mathbf{r'}m\sigma}^{\vphantom{dagger}}+c^{\dagger}_{\mathbf{r'}m\sigma}c_{\mathbf{r}m\sigma}^{\vphantom{dagger}}) \\
&-V \sum\limits_{\mathbf{r}\sigma}
(c^{\dagger}_{\mathbf{r}1\sigma}c_{\mathbf{r}2\sigma}^{\vphantom{dagger}}+c^{\dagger}_{\mathbf{r}2\sigma}c_{\mathbf{r}1\sigma}^{\vphantom{dagger}}) \\
&+ U \sum\limits_{\mathbf{r}m} (n_{\mathbf{r}m\uparrow}-\frac{1}{2})
(n_{\mathbf{r}m\downarrow}-\frac{1}{2}) \\
&+\sum\limits_{\mathbf{r}m\sigma}  ( \Delta e^{i\theta_{m}}  * (-1)^{x+y} - \mu)  n_{\mathbf{r}m\sigma}
\end{split}
\label{H}
\end{equation}
where $m=1,2$ labels two layers (orbitals) while ${\bf r,r'}$ are site indices and $\sigma$ denotes spin.  The first two terms are intra- and inter-layer nearest-neighbor hopping.  We consider a square lattice with intralayer hopping $t=1$ setting the energy scale. The interlayer (interorbital) hybridization $V$ and on-site repulsion $U$ are two parameters for conventional bilayer Hubbard model. The repulsive on-site interaction term is written in particle-hole symmetric form so that at $\mu=0$ the system is half-filled. The last term denotes the staggered (expressed as $(-1)^{x+y}$) potential for two layers. Note that in general this potential can have an additional phase degree of freedom $\theta_{m}$ while in this paper we only considers two special cases $\theta_{2}=0$ (in-phase) and $\theta_{2}=\pi$ (anti-phase) with fixed $\theta_{1}=0$.

In the absence of the external staggered potential, the noninteracting limit $U=0$ has two bands for each spin,
\begin{equation}
\epsilon_{\mathbf{k} \sigma}^{\pm} = - 2t(\cos k_{x}+\cos k_{y}) \pm V
\label{nonintbands}
\end{equation}
so that $V/t\leq 4$ yields metallic behavior while $V/t>4$ characterizes a band insulator (BI) with gap $2(V - 4t)$ splitting bonding and antibonding bands. The phase diagram at finite interaction $U$ is still in debate although the model has been studied extensively using different numerical methods such as determinant quantum Monte Carlo (DQMC)\cite{bilayerDQMC}, dynamical mean field theory (DMFT)\cite{bilayerDMFT}, variational Monte Carlo (VMC)\cite{bilayerVMC}, and functional renormalization group (fRG)\cite{bilayerFRG}. It is generally agreed that large $U$ leads to a direct transition from a Neel phase to a singlet as the interlayer hopping $V$ is increased. However, the properties at small $U$ was controversial since DQMC\cite{bilayerDQMC} and DMFT\cite{bilayerDMFT} studies suggest a paramagnetic metallic intermediate phase while VMC\cite{bilayerVMC} study predicts a direct transition as for large $U$. Recent work employing unbiased functional renormalization group approach demonstrated that any weak finite $U$ would induce an antiferromagnetic Mott-insulator\cite{bilayerFRG} . They further resolved the difficulty of DQMC for identifying this antiferromagnetic ground state for finite interlayer hopping in the weak-coupling regime, where nonmonotonic finite-size corrections are related to the two-sheeted Fermi surface structure of the metallic phase.

In the absence of interlayer hybridization, namely $V/t=0$, incorporating a staggered potential in one-band tight-binding models on a bipartite lattice with band $\epsilon_{\mathbf{k}}$ mixes momentum states ${\bf k}$ and ${\bf k + \pi}$ to open up a spectral gap with dispersion relation $E_{\mathbf{k}} = \pm \sqrt{\epsilon_{\mathbf{k}}^2 + \Delta^2}$. Such a staggered potential couples strongly to charge density wave since it provides a one-body energy which favors an oscillating charge density on the two sublattices.

\begin{figure}
\psfig{figure=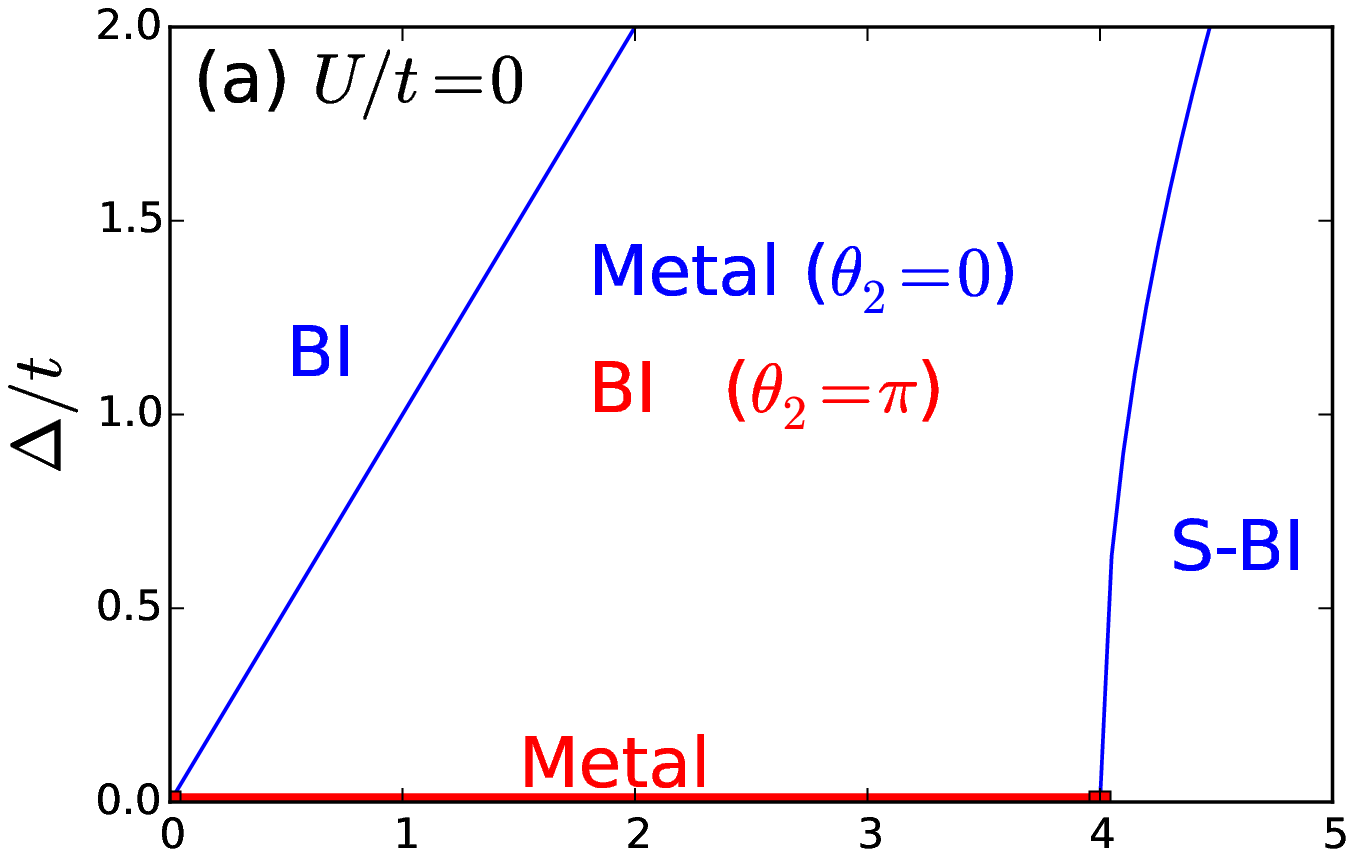,height=5.0cm,width=8.8cm,angle=0,clip,trim={0 0.8cm 0 0}}
\psfig{figure=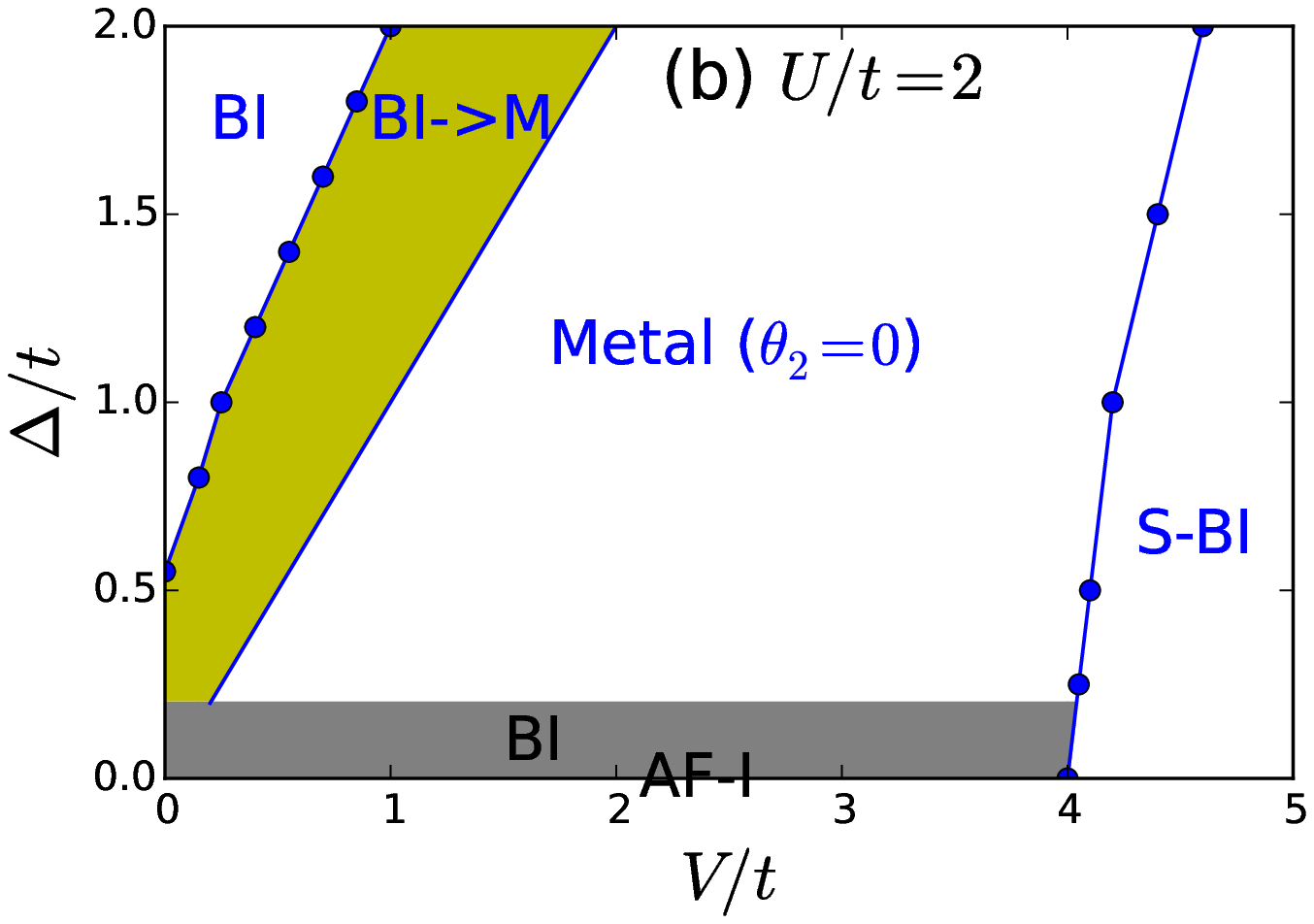,height=5.1cm,width=8.8cm,angle=0,clip,trim={0 0 0 0.7cm}}
\caption{(Color online) Phase diagrams of bilayer ionic Hubbard model at lowest accessible temperature $\beta t=20$ for (a) $U/t=0$ and (b) $U/t=2$ determined via spectral functions and temperature dependence of local Green's functions. Blue (red) color is to differentiate two types of staggered potentials: $\theta_{2}=0$ in-phase and $\theta_{2}=\pi$ anti-phase. Only systems with $\theta_{2}=0$ can host interaction-driven insulator-metal transition (yellow patch). The gray patched regime is unresolved ultimately due to unaffordable larger lattice sizes and lower temperatures. BI, S-BI, and AF-I denote band insulator, singlet insulator, and antiferromagnetic insulator respectively. The system size is $16 \times 16 \times 2$.}
\label{phase}
\end{figure}

For both finite $\Delta$ and $V$, we first take a glance at the noninteracting case, namely the $U=0$ limit of Eq.\ref{H}, at which the hamiltonian can be transformed to be
\begin{equation}
\begin{split}
          \left(
            \begin{array}{l}
              c^{\dagger}_{1\mathbf{k}} \\
              c^{\dagger}_{1\mathbf{k}+\pi} \\
              c^{\dagger}_{2\mathbf{k}} \\
              c^{\dagger}_{2\mathbf{k}+\pi} \\
            \end{array}
          \right)^{T}
          \left(
            \begin{array}{cccc}
              \epsilon_{\mathbf{k}}  & \Delta e^{i\theta_{1}} & V & 0 \\
              \Delta e^{i\theta_{1}} & -\epsilon_{\mathbf{k}} & 0 & V \\
              V & 0 &  \epsilon_{\mathbf{k}}  & \Delta e^{i\theta_{2}} \\
              0 & V &  \Delta e^{i\theta_{2}} & -\epsilon_{\mathbf{k}}  \\
            \end{array}
          \right)
          \left(
            \begin{array}{l}
              c^{\phantom{\dagger}}_{1\mathbf{k}} \\
              c^{\phantom{\dagger}}_{1\mathbf{k}+\pi} \\
              c^{\phantom{\dagger}}_{2\mathbf{k}} \\
              c^{\phantom{\dagger}}_{2\mathbf{k}+\pi} \\
            \end{array}
          \right)
\end{split}
\label{Hk}
\end{equation}
whose general four-band structure is complicated but we have two special cases: $E^{2}_{\mathbf{k}} = ( \sqrt{\epsilon_{\mathbf{k}}^2 + \Delta^2} \pm V)^{2}$ for $\theta_{2}=0$ and $E^{2}_{\mathbf{k}} = (\epsilon_{\mathbf{k}} \pm V)^{2} + \Delta^{2}$ for $\theta_{2}=\pi$ with fixed $\theta_{1}=0$, whose corresponding phase diagrams at half-filling are given in Fig.~\ref{phase}(a). Although the case of $\theta_{2}=\pi$ is plain since the system transits to an insulator immediately after turning on the staggered potential, the case of $\theta_{2}=0$ shows a richer phase diagram. In particular, different from the conventional Hubbard bilayer, there is an additional V-driven BI-Metal transition at $V=\Delta$. Besides, the alternating site energies induced by $\Delta$ smears the bandwidth from $4t$ to $\sqrt{(4t)^{2}+\Delta^{2}}$ so that the critical $V$ for Metal-Singlet transition increases with $\Delta$.

From now on we concentrate on the system in the presence of intermediate Hubbard interaction $U$ that can host an intermediate metallic\cite{comment} phase in single-layer IHM.
We treat the interaction term in Eq.\ref{H} using determinant quantum Monte Carlo (DQMC) technique\cite{blankenbecler81}, which is numerically exact in principle to solve interacting tight binding electron Hamiltonians. Although DQMC has the advantage of being able to incorporate and measure magnetic, charge, and pairing correlations between spatial sites, it is formulated on finite lattices so that finite-size effects can be significant and must be assessed carefully. Since we concentrate on the impact of interlayer hybridization on the single-layer IHM, whose intermediate metallic phase merely occupies a narrow region at weak interaction $U$ in its phase diagram\cite{RTS2007,ionic1,QMC}, we focus on $U/t=2.0$. Another practical reason for this choice lies that larger $U$ leads to severe sign problem even for relatively small lattices, e.g. $12\times 12$ for $\theta_{2}=0$ due to the lack of bipartition. Obviously, there is no sign problem for $\theta_{2}=\pi$ by the particle-hole symmetry in bipartite lattices.
Nevertheless, small $U$ imposes strong requirements for large enough lattice sizes and low enough temperatures to see the insulating behavior at half-filling\cite{U2DQMC}. Therefore, most of the results presented in this paper will be for two $16\times 16$ layers to alleviate the finite-size effects as best as possible. Besides, most of our simulations are based on the inverse temperature $\beta t=20$, except for some cases with severe sign problems, where we adopt $\beta t=14$ and for temperature-dependence of local Green's functions, where the lowest accessible temperature is $\beta t=25$ for $14\times 14\times 2$ bilayer. However, we observed that in most cases $\beta t=14$ is low enough to qualitatively catch the basic physical properties illustrated in this paper.

We quantify the effects of hybridization $V$ by the spectral functions and the antiferromagnetic structure factors of charge/spin density waves. Our main result is the phase diagram Fig.~\ref{phase}(b), whose phase boundaries are mainly obtained via the spectral functions combined with both the temperature-dependence of local Green's functions and finite-size scaling arguments. We use blue (red) color to differentiate two types of staggered potentials.
We argue that only the systems with in-phase potential $\theta_{2}=0$ can host the interaction-driven BI-M transition.
The most distinctive feature is the enlargement of the metallic regime (yellow patched) compared with the conventional 2D IHM. Precisely, the interaction-driven intermediate metallic phase occurs not only for single-layer IHM, namely $V/t=0$, but also in the presence of interlayer hybridization. However, anti-phase ionic potential always induces an insulator with charge density order due to the coupling of potential peak and valley between layers. Regarding the robustness of the intermediate metallic phase in 2D IHM, they are immune to the interlayer hybridization for in-phase potential; while fragile tending to the insulating behavior for anti-phase potential.

Concerning weak $\Delta/t$, we should point out that our DQMC simulations cannot conclusively capture the possible spectral gap and/or long range magnetic order to distinguish the phases due to the unaffordable larger lattice sizes and lower temperatures. At $\Delta/t=0$, we follow the claiming of previous functional renormalization group (fRG) study\cite{bilayerFRG} that any finite Hubbard interaction generates an insulator for $V/t\lesssim 4$ at half-filling. Via the temperature dependence of local Green's functions, we believe that this insulating phase extends to finite $\Delta$ region (gray patched) not only in 2D ionic model\cite{RTS2007} but also for finite $V$. The exact phase boundary for this regime in the thermodynamic limit deserves further study employing other appropriate methods.

\section{Spectral functions}

We first discuss how we obtained the phase diagram in Fig.~\ref{phase} via the spectral properties. In order to quantify it to distinguish metals from insulators, we examined the single-particle local density of states (DOS), which is obtained by an analytic continuation of the local imaginary-time dependent Green's function
$G_{\pm}(\tau)=- \sum\limits_{{\bf j}} \langle
c^{\phantom{\dagger}}_{{\bf j}\pm}(\tau) c^{\dagger}_{{\bf
j}\pm}(0) \rangle$ by inverting
\begin{equation}
   G_{\pm}(\tau)= \int_{-\infty}^{\infty}\ d\omega \frac{e^{-\omega\tau}}{e^{-\beta\omega}+1}\ N_{\pm}(\omega) \label{aw}
\end{equation}
using the maximum entropy method\cite{gubernatis91}. To avoid the ambiguity from analytical continuation of Green's functions, we provide the original data of local $G(\tau)$ in the appendix as well, where we show that both local $G(\tau)$ and associated $N(\omega)$ are consistent with each other. Note that the spin indices are omitted due to the spin symmetry while $\pm$ denotes two inequivalent lattice sites with potential amplitude $\pm \Delta$ respectively. We will only show the results of $N_{-}(\omega)$ through which $N_{+}(\omega)=N_{-}(-\omega)$. Our focus will be on DOS at the Fermi surface $N_\sigma(\omega=0)$ to determine the phase.

Intuitively, on the one hand, the system with $\theta_{2}=0$ in-phase potential has inversion symmetry between two layers, whose behavior should be similar to the conventional Hubbard bilayers for $\Delta=0$, where the hybridization $V$ simply pushes the system towards metals regardless of the original phase at $V/t=0$.
On the other hand, $\theta_{2}=\pi$ (anti-phase) potential breaks the inversion symmetry and couple one layer's potential valley to the other layer's potential peak so that the electrons with opposite spins can occupy the potential valley in two layers alternatively, which brings the system into an insulator with strong charge density order.

\begin{figure}
\psfig{figure=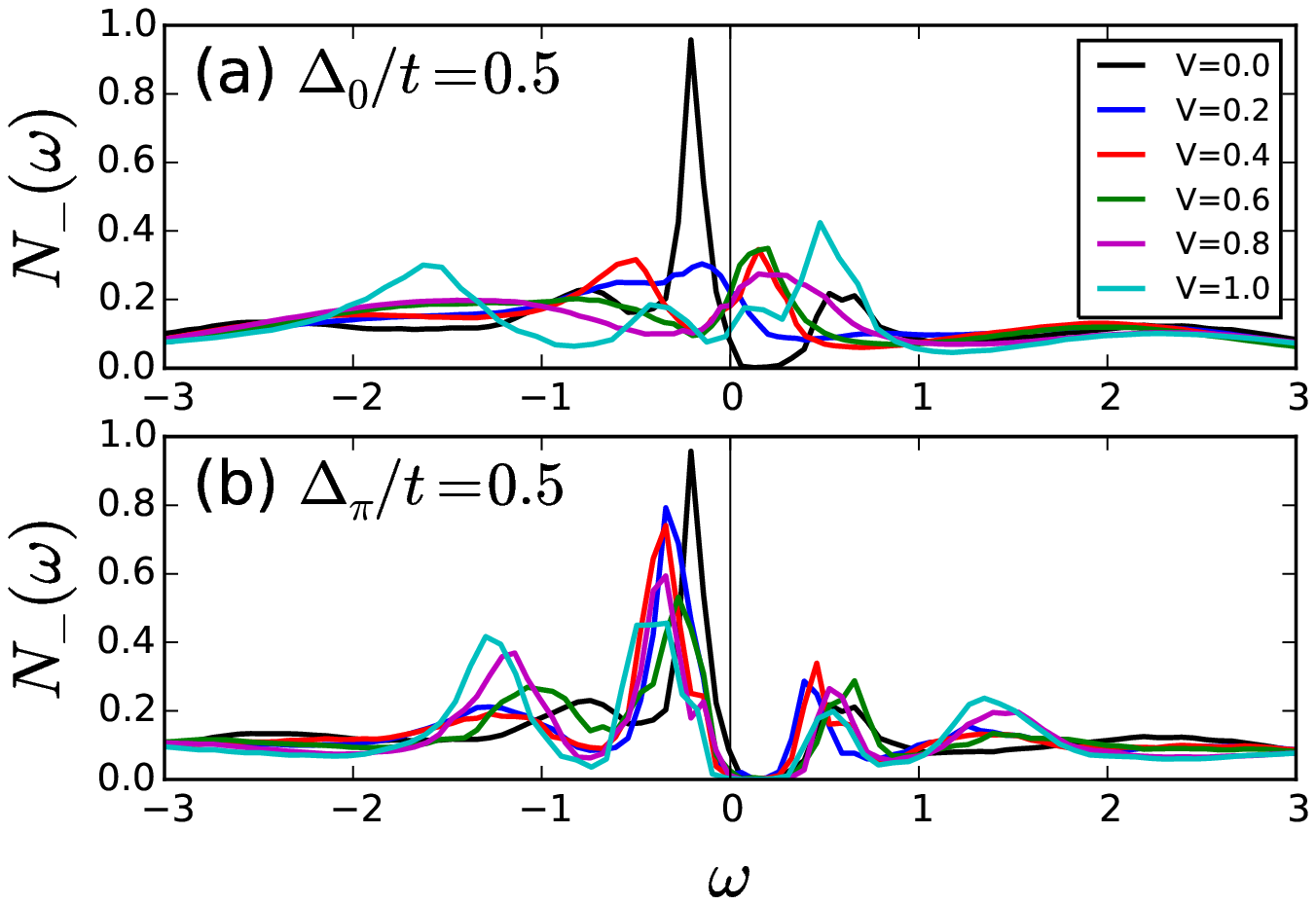,height=5.6cm,width=8.8cm,angle=0,clip,trim={0 0.8cm 0 0}}
\psfig{figure=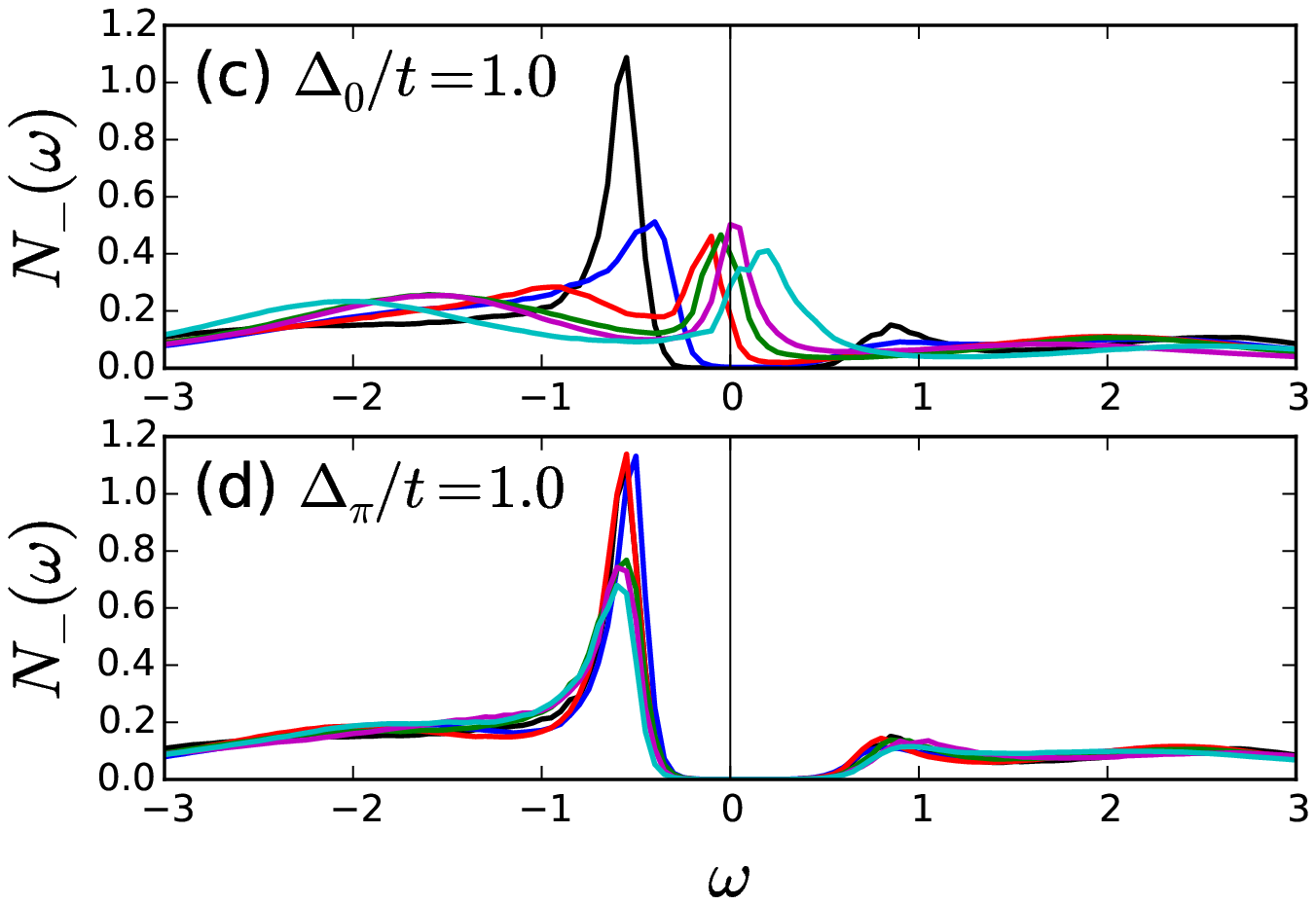,height=5.6cm,width=8.8cm,angle=0,clip,trim={0 0 0 0.7cm}}
\caption{(Color online) Comparison of local density of states for potential valley site with varying interlayer hybridization $V$ starting from metallic ($\Delta/t=0.5$) and insulating ($\Delta/t=1.0$) phases in both cases of $\theta_{2}=0$ in-phase and $\theta_{2}=\pi$ anti-phase staggered potentials. The hybridization pushes the in-phase (anti-phase) system towards a metal (insulator). $\Delta_{0}$ and $\Delta_{\pi}$ denote $\theta_{2}=0$ and $\theta_{2}=\pi$ separately for simplicity.}
\label{NwsmallV}
\end{figure}

Fig.~\ref{NwsmallV} illustrates the local density of states for potential valley sites, which generally confirms the expectation discussed above. (Note that we use $\Delta_{0}$ and $\Delta_{\pi}$ to denote the cases of $\theta_{2}=0$ and $\theta_{2}=\pi$ separately for simplicity.) $\Delta/t=0.5$ and $\Delta/t=1.0$ correspond to metallic and insulating phases in single-layer IHM, respectively.
The general trend towards a metallic phase for $\theta_{2}=0$ can be seen in panels (a) and (c) so that there exists an V-driven insulator-metal transition for $\Delta/t=1.0$. Obviously, the hybridization does not qualitatively modify the original single-layer metallic behavior at $\Delta/t=0.5$.
Increasing $V$ only smoothly refines the electronic four-band structure. In particular, turning on $V$ induces the initial shrink and subsequent right shift of the spectral peak. Regarding the metallic behavior at relatively large $V>\Delta$, the system can be imagined as an interlayer dimer moving around in a square lattice. Moreover, similar to Fig.~\ref{phase}(a), the phase boundary (blue dotted line) in Fig.~\ref{phase}(b) indicates the competition between two energy scales $\Delta$ and $V$. Apparently larger $\Delta$ requires stronger hybridization $V$ to release the potential energy of electrons to accomplish the insulator-metal transition.

In contrast, panels (b) and (d) show distinct behavior for anti-phase staggered potential which couples the potential peaks and valleys between layers. In panel (b), the original exotic in-plane metallic behavior in single-layer IHM is quickly weakened by the perpendicular hybridization, which is accompanied with gradual spectral weight redistribution towards higher energies for the spectral gap opening. Further evidence on the robust trend to insulators is illustrated in panel (d), which only involves the peak shrink with increasing $V$ even for relatively large hybridization.
As expected intuitively, the insulating behavior should also be characterized by the double occupancy of electrons with opposite spins at potential valley alternatively between layers, which induces strong in-plane charge density wave but weak spin density wave (more details in the next section). In fact, the lack of this alternative potential profile also explains the absence of metal-insulator transition at small $V$ for $\theta_{2}=0$.

\begin{figure}
\psfig{figure=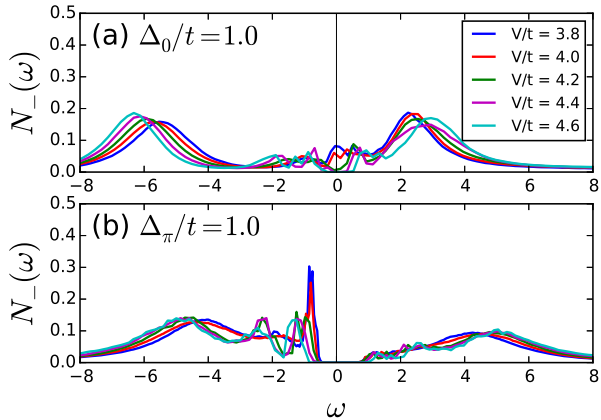,height=6.0cm,width=8.8cm,angle=0,clip}
\caption{(Color online) Comparison of local density of states for potential valley site with large interlayer hybridization $V\gg \Delta$. The splitting between the bonding and antibonding bands induced by $V$ results in the formation of tightly bound singlets. The symmetry/asymmetry of the two band peaks indicates different potential environment imposed on the singlets between in-phase (a) and anti-phase (b) cases.}
\label{NwlargeV}
\end{figure}

The above discussion applies for small to intermediate ratio $V/\Delta$. At large enough hybridization $V\gg \Delta$, the effects of both in-phase and anti-phase staggered potential can be neglected. The splitting between the bonding and antibonding bands induced by $V$ results in spectral gap opening and a transition to the singlet phase that is reminiscent of the conventional bilayer Hubbard model. Fig.~\ref{NwlargeV} compares the spectral properties for two cases, both of which clearly show the formation of bonding and antibonding bands with the peak separation around $2V$ and panel (a) provides the evolution of spectral behavior across the metal-singlet transition for $\theta_{2}=0$. One important feature at large $V$ concerns the positions of bonding and antibonding band peaks. The peak position asymmetry in panel (a) can be attributed as the asymmetric energy shift induced by the staggered potential on the tightly bound singlets with apparent lower energy in potential valleys; while the symmetric peak distribution in panel (b) reflect the potential peak-valley coupling between layers which provides more homogeneous potential on the singlets. Besides, in panel (b) the remaining spectral peak near the gap edge, although largely shrinks compared with that for weaker hybridization, provides more evidence of the robustness on $V$ for the anti-phase staggered potential. In fact, this additional peak corresponds to the charge density order discussed below.

\section{Temperature dependence of local $G(\tau)$}

Since our DQMC simulations are performed only for finite temperature, it is desirable to explore the temperature evolution of the interaction-driven metallic phase to investigate if it is the ultimate ground state. To this aim, we adopt the approximate formula relating the density of state (DOS) at Fermi energy and the local Green's function to avoid the intrinsic ambiguity of the analytic continuation within the maximum entropy method
\begin{equation}
  N(0) \approx -\beta G(\mathbf{r}=0, \tau=\beta/2)/\pi  \label{Gtau}
\end{equation}
which involves the assumption that the temperature is much lower than the energy scale $\Omega$ on which there are structures in DOS\cite{GtauNw}.

\begin{figure}
\psfig{figure=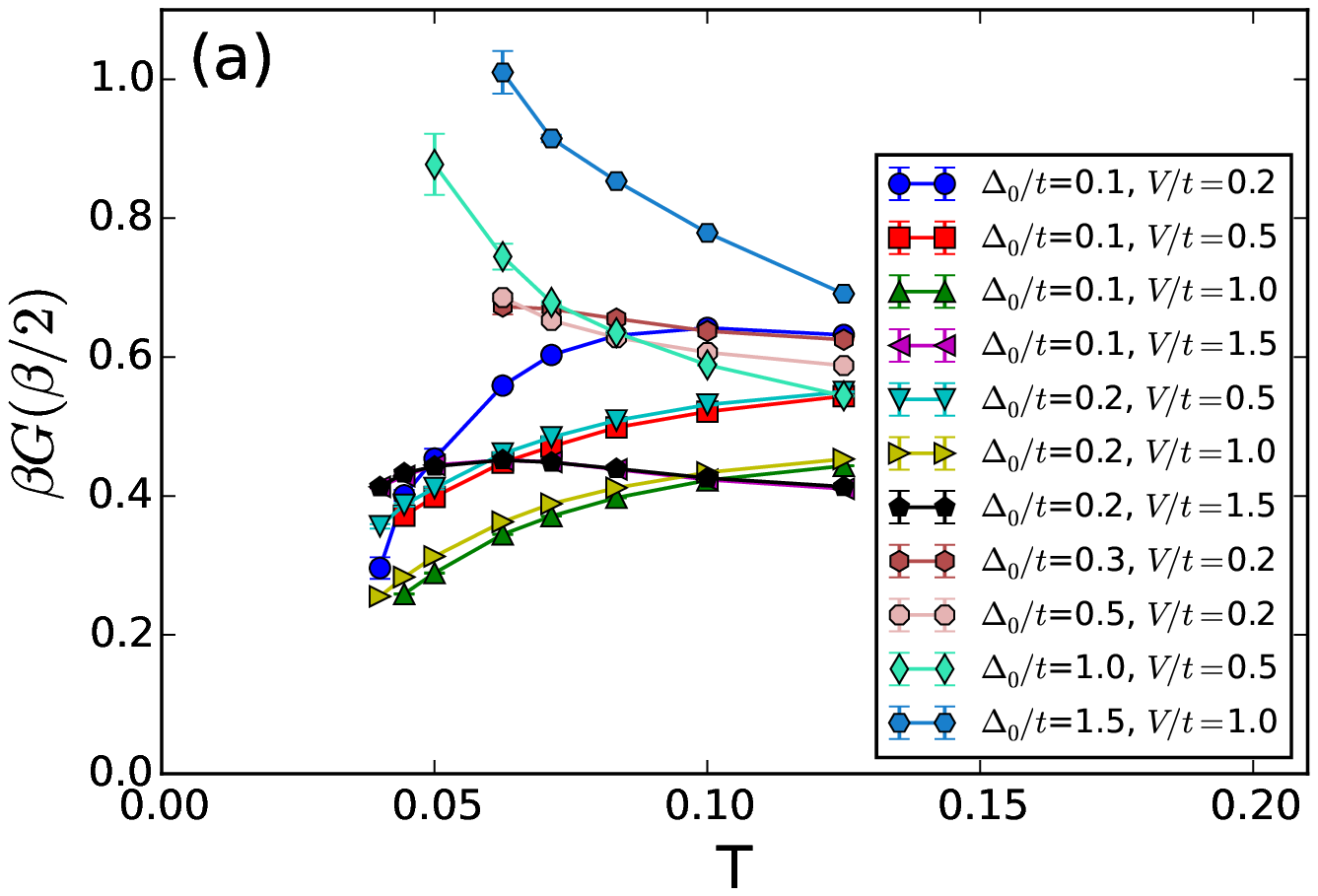,height=5.2cm,width=8.8cm,angle=0,clip,trim={0 0.8cm 0 0}}
\psfig{figure=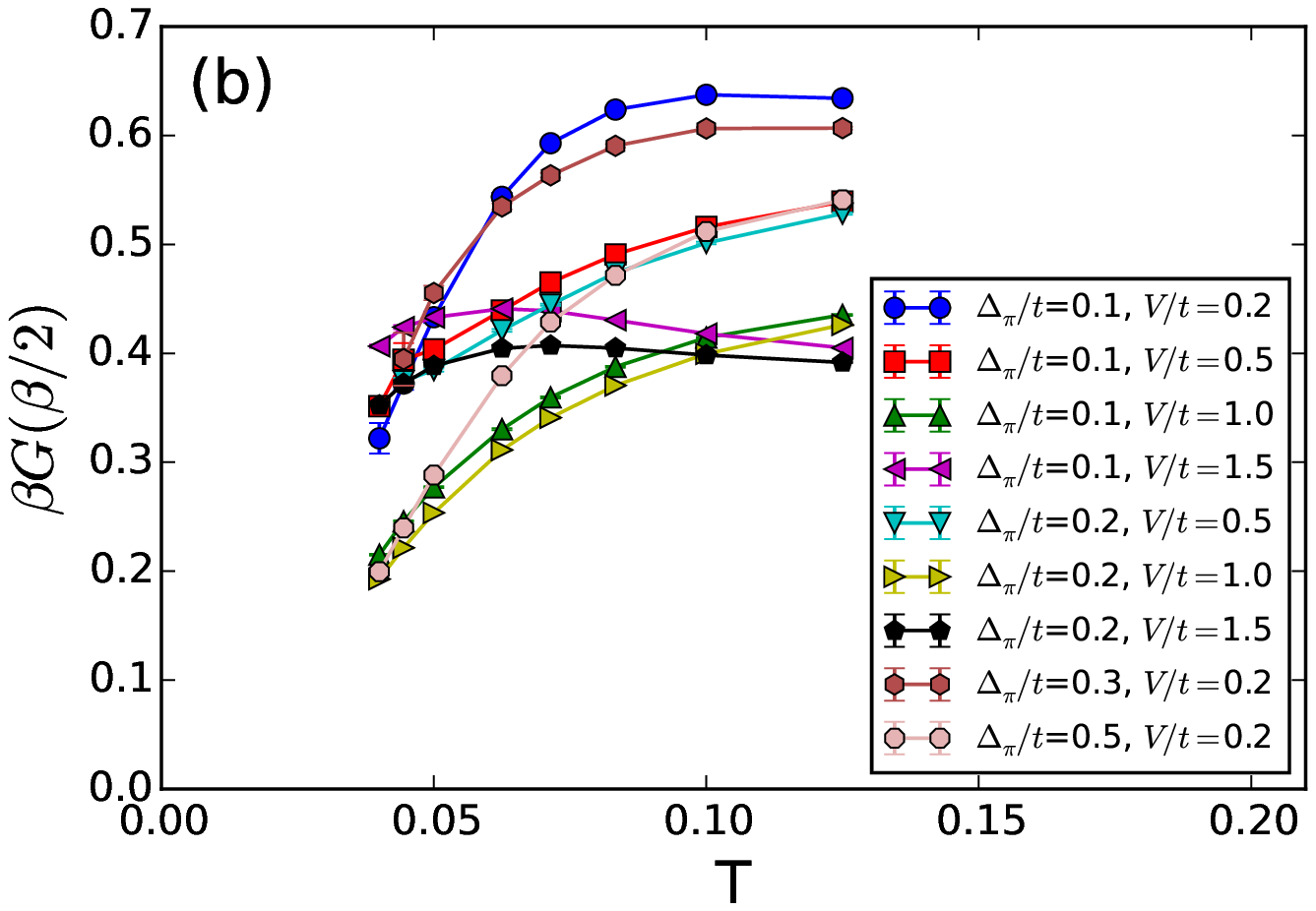,height=5.2cm,width=8.8cm,angle=0,clip,trim={0 0 0 0.7cm}}
\caption{(Color online) Temperature evolution of local $\beta G(\beta/2)$ at potential valley sites for in-phase and anti-phase ionic potentials for bilayer $N=14 \times 14 \times 2$. At weak $\Delta$, $\beta G(\beta/2)$ decreases with lowering temperature implying the insulating nature of both systems. At larger $\Delta/t \geq 0.3$, only in-phase ionic potential can host the interaction-driven metallic phase, although the available data is limited by the sign problem for unaffordable larger lattice sizes and lower temperatures.}
\label{GtauT}
\end{figure}

At $\Delta/t=0$, namely the conventional bilayer Hubbard model, the previous functional renormalization group (fRG) study\cite{bilayerFRG} pointed out that any finite Hubbard interaction generates an antiferromagnetic insulator (AF-I) for $V/t\lesssim 4$ at half-filling. They further resolved the difficulty of DQMC for identifying this AF-I ground state for finite interlayer hopping in the weak-coupling regime. Therefore, it is natural to expect that this AF-I would survive even for turning on weak $\Delta$. As illustrated in Fig.~\ref{GtauT}, for both cases of (a) in-phase and (b) anti-phase ionic potentials, $\beta G(\beta/2)$ decreases with lowering temperature at weak $\Delta$, which implies the insulating nature of these systems.
However, for larger $\Delta/t  \gtrsim 0.3$, they induce different behavior of $\beta G(\beta/2)$ vs $T$, which provides strong evidence that only in-phase ionic potential can host the interaction-driven metallic phase. Although our DQMC simulations cannot conclusively distinguish the metallic or insulating phase at $T=0$ due to the unaffordable larger lattice sizes and lower temperatures so that we denote the unresolved regime of weak $\Delta/t \leq 0.2$ by gray patch in Fig.~\ref{phase}(b), we believe that the distinctive increase of $\beta G(\beta/2)$ with $T$ for $\Delta_{0}/t=1.0, V/t=0.5$ and $\Delta_{0}/t=1.5, V/t=1.0$ still provides stimulating insights on the lower temperature properties. The ultimate ground states for both interaction-driven metallic phase and weak $\Delta$ regime in the thermodynamic limit deserves further study employing other appropriate methods.

\section{Structure factors}

With the above knowledge of local Green's functions and spectral properties, we want to further understand the competition between staggered potential and interlayer hybridization by exploring the structure factors of charge/spin density waves. The staggered potential imposes a checkerboard modulation on the electron density in each layer so that the antiferromagnetic charge and/or magnetic correlations are expected and should be characterized by the corresponding structure factors with ordering wave vector $\mathbf{k}_{\pi} = (\pi, \pi)$.

We measure the in-plane charge and spin density wave structure factors
\begin{equation}
  S^{CDW}_{SDW} = \frac{1}{N} \sum\limits_{\bf l,j}
e^{i\mathbf{k}_{\pi}\cdot (\mathbf{l}-\mathbf{j})} \langle
(n_{{\bf l}\uparrow} \pm n_{{\bf l}\downarrow})
(n_{{\bf j}\uparrow} \pm n_{{\bf j}\downarrow}) \rangle
\end{equation}
and investigate their evolution with the hybridization $V$.

\begin{figure}
\psfig{figure=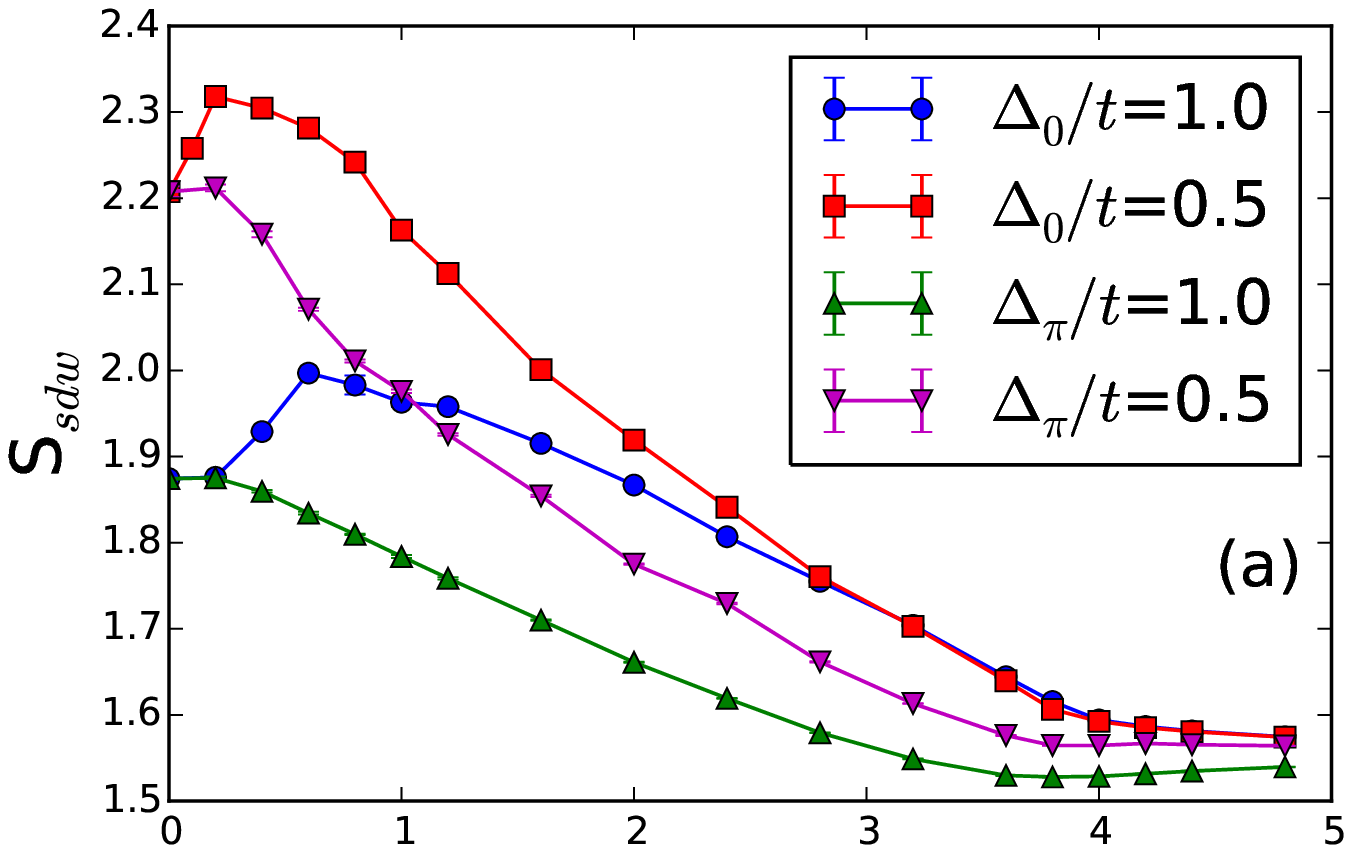,height=5.2cm,width=8.8cm,angle=0,clip,trim={0 0.8cm 0 0}}
\psfig{figure=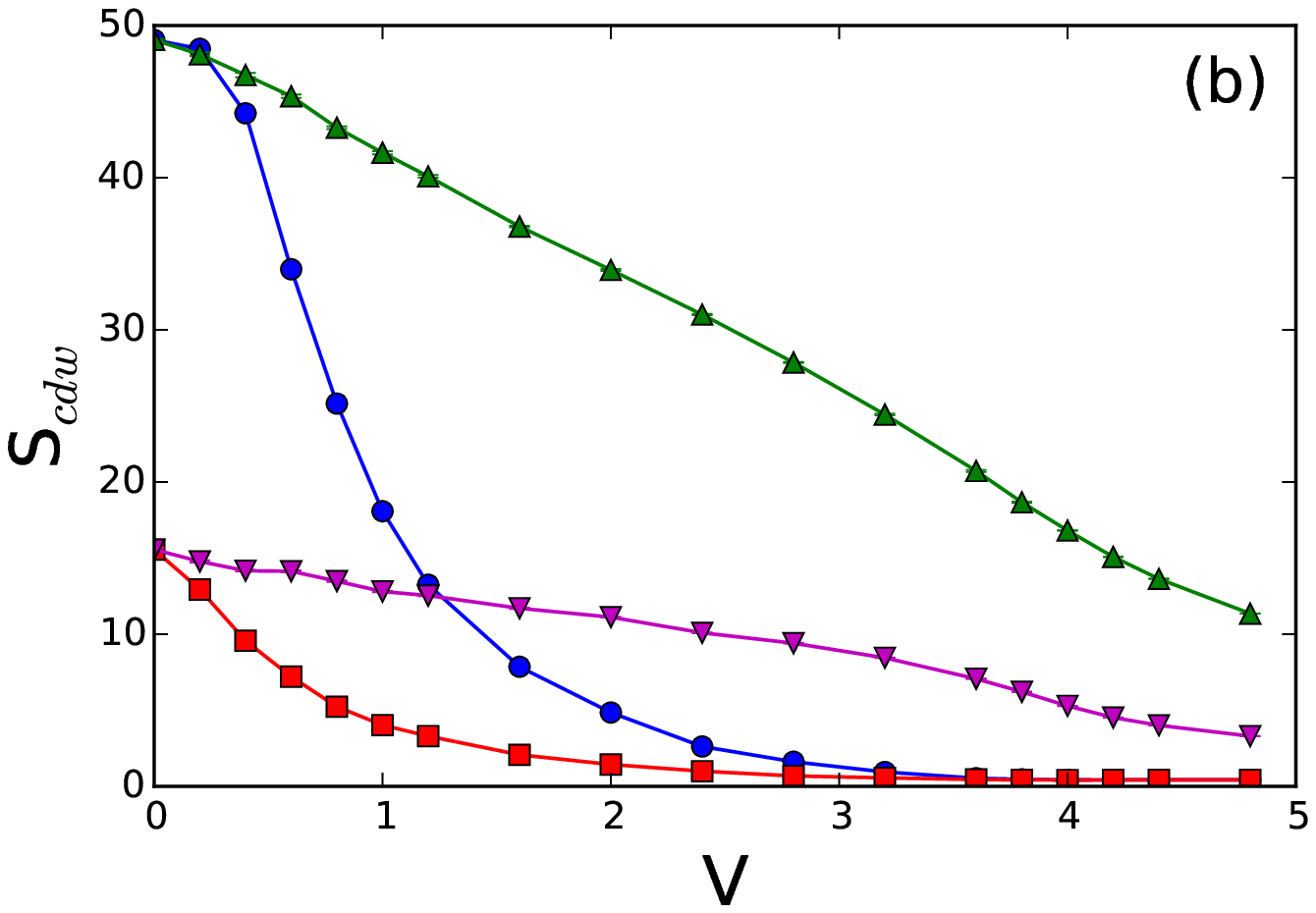,height=5.2cm,width=8.8cm,angle=0,clip,trim={0 0 0 0.7cm}}
\caption{(Color online) Antiferromagnetic structure factor of (a) spin density wave $S_{sdw}$ and (b) charge density wave $S_{cdw}$ with increasing interlayer hybridization $V$ for both in-phase and anti-phase staggered potential systems.}
\label{SDWCDW}
\end{figure}

As shown in Fig.~\ref{SDWCDW}, there is no doubt that the general trend for both $S_{sdw}$ and $S_{cdw}$ is decreasing with $V$ due to the formation of tightly bound singlets at large enough $V$. In addition, stronger staggered potential (larger $\Delta$) leads to stronger antiferromagnetic charge correlations but weaker magnetic correlations. Panel (a) demonstrates that systems with $\theta_{2}=0$ in-phase staggered potential have more spin degrees of freedom so that larger $S_{sdw}$ due to the checkerboard profile of bilayer potential compared with the case of $\theta_{2}=\pi$, which instead promotes strong antiferromagnetic charge correlation as shown in panel (b). As discussed previously, the potential peak-valley coupling between layers for $\theta_{2}=\pi$ preserves the alternating potential profile so that $V$ does not harm the charge density order as much as for $\theta_{2}=0$.

Moreover, Fig.~\ref{SDWCDW} shows one distinct feature of $S_{sdw}$ for $\theta_{2}=0$, which has a peak at $V/t=0.6$ ($V/t=0.2$) for $\Delta_{0}/t=1.0$ ($\Delta_{0}/t=0.5$). This nontrivial phenomena can be traced back to its spectral properties. As shown in Fig.~\ref{NwsmallV}(a) and (c), the increase of $S_{sdw}$ is accompanied with the insulator-metal transition and continued within the metallic phase. Although this peak structure probably only occurs in finite size systems as discussed in next section, we can imagine that the interlayer hybridization releases some electronic kinetic energies locked by the in-phase staggered potential so that promotes the magnetic correlation. Certainly the magnetic correlation is unstable and will be dominated by the metallic nature of the system so that decreases with $V$ finally.

\section{Finite-size effects}

\begin{figure}
\psfig{figure=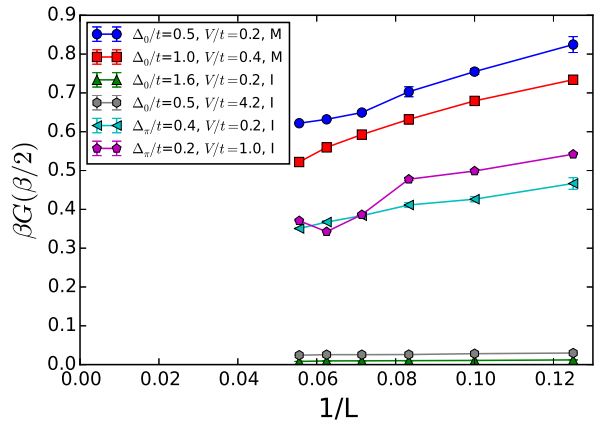,height=6.0cm,width=8.8cm,angle=0,clip}
\caption{(Color online) The local $\beta G(\beta/2)$'s dependence on lattice sizes for metals and insulators determined in $N=16\times 16 \times 2$ systems. For in-phase potential, the insulators do not show finite-size effects while metals show noticeable effects that do not change the phase qualitatively. The two curves for anti-phase potential are only for comparison since Fig.~\ref{GtauT} shows that this type of potential cannot host metallic phases.}
\label{G0finitesize}
\end{figure}

\begin{figure}
\psfig{figure=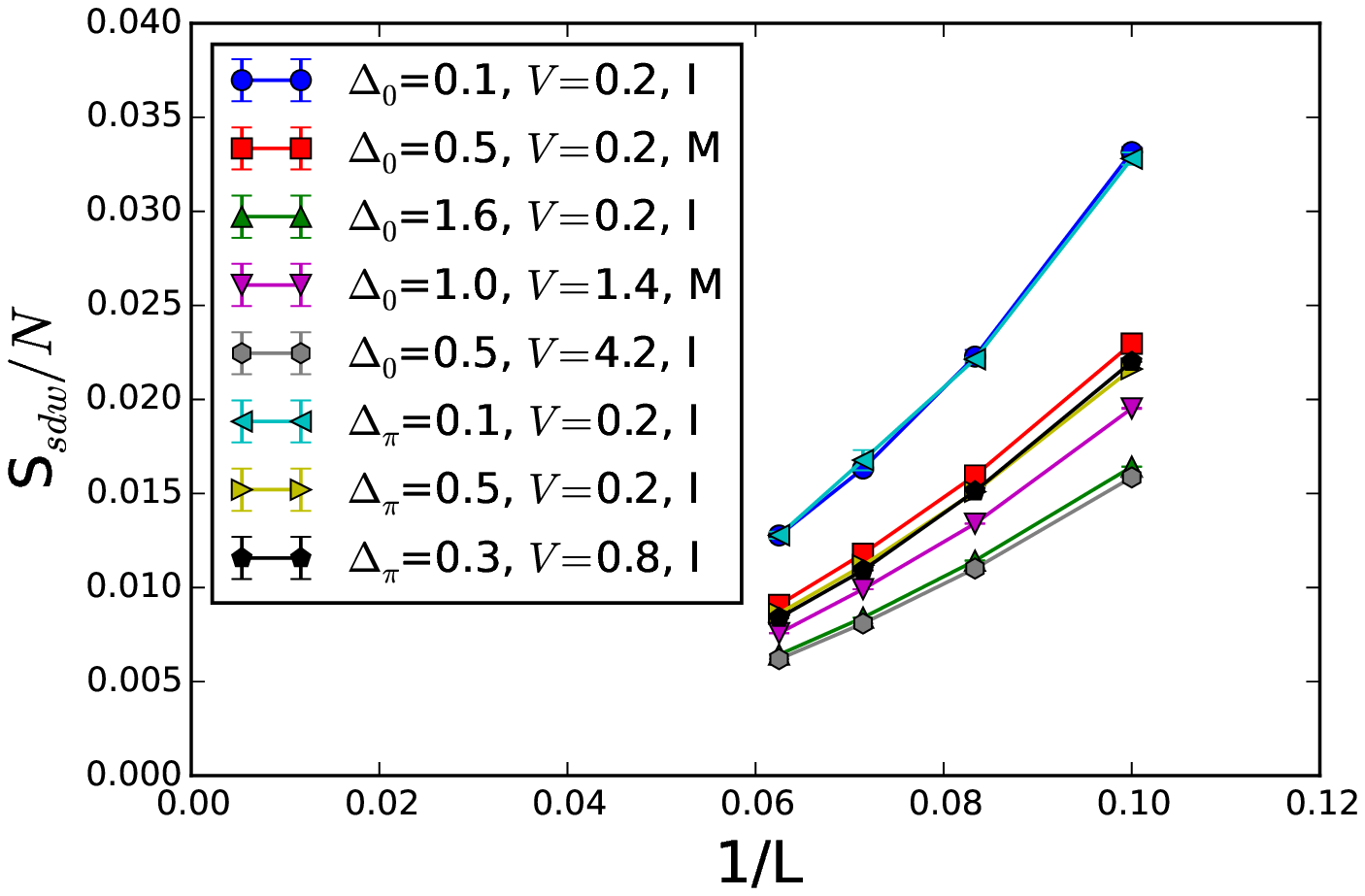,height=5.2cm,width=8.8cm,angle=0,clip,trim={0 0.8cm 0 0}}
\psfig{figure=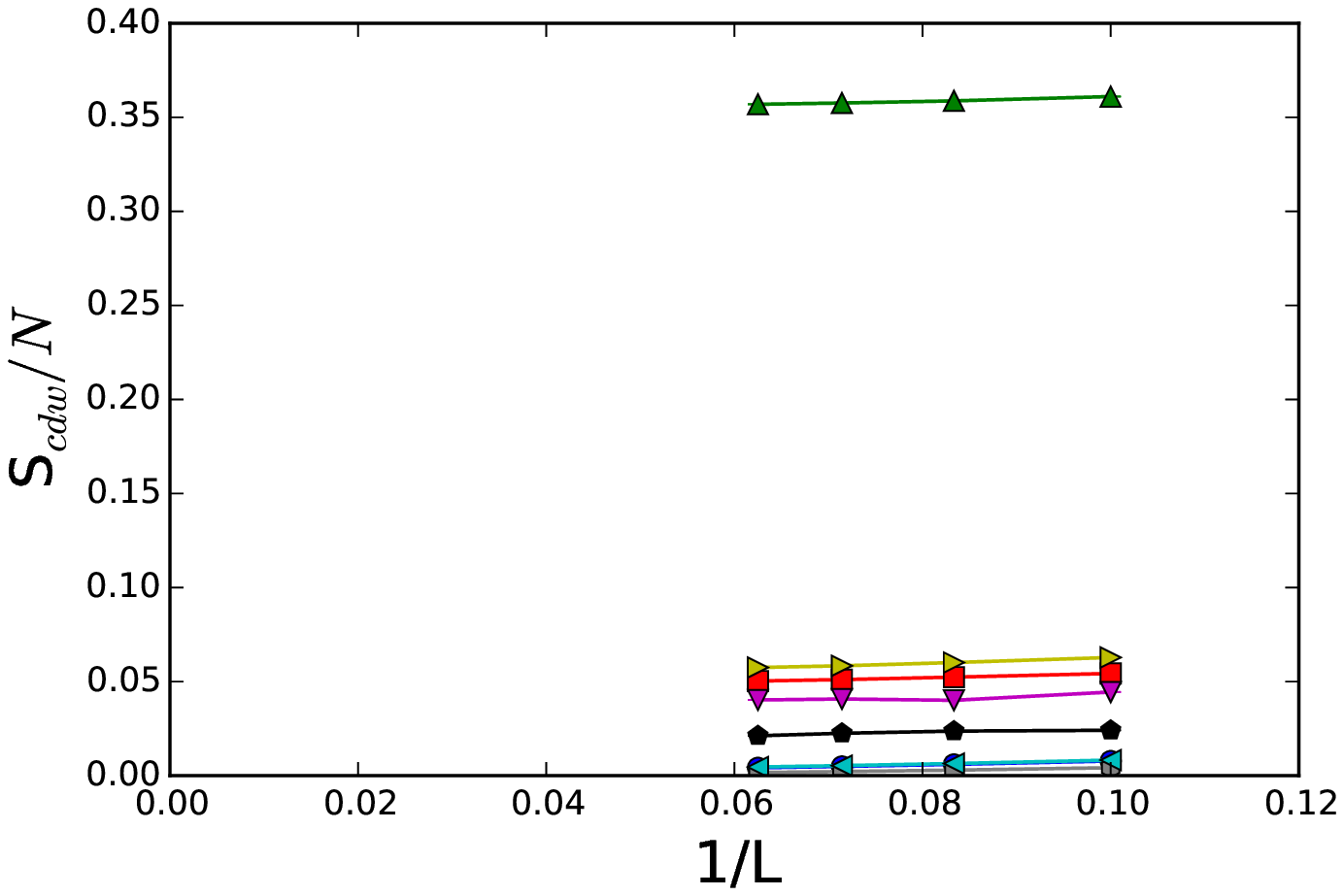,height=5.2cm,width=8.8cm,angle=0,clip,trim={0 0 0 0.7cm}}
\caption{(Color online) Finite-size effects of the antiferromagnetic structure factors. Strong finite-size effect of $S_{sdw}/N$ indicates the absence of antiferromagnetic spin order in the thermodynamic systems; while the lack of $S_{cdw}/N$'s size dependence confirms the charge density order due to the staggered potential.}
\label{SDWfinitesize}
\end{figure}

As mentioned before, one important issue associated with small $U/t=2.0$ is its large finite-size effects and normally large enough lattice sizes and low enough temperatures are required to capture the insulating behavior. Therefore, although the results shown in this paper are for the largest affordable size $16\times 16 \times 2$, it is important to discuss more details on the finite-size effects of different physical properties.

To avoid the intrinsic ambiguity of maximum entropy method, here we again adopt the approximate formula Eq.~\ref{Gtau} to investigate the finite-size effects of spectral properties.
Fig.~\ref{G0finitesize} provides some flavors on the behavior of local $\beta G(\beta/2)$ in different lattice sizes for both metal (M) and insulator (I) determined in $N=16\times 16 \times 2$ lattices. Clearly, the insulators for $\Delta_{0}/t=1.6, V/t=0.2$ and $\Delta_{0}/t=0.5, V/t=4.2$ have no finite-size effects. Nevertheless, the interaction-driven metallic phase for $\theta_{2}=0$ in-phase potential shows noticeable dependence on the lattice size, which does not signal a qualitative change of the metallic nature. Although we believe that the anti-phase ionic potential always induces an insulator as discussed in Sec. IV, we show their finite-size effects for comparison, which has similar decrease with larger lattice size as in-phase potential. As evidenced in Fig.~\ref{GtauT}, the local $\beta G(\beta/2)$ would gradually decrease to open a spectral gap.

In addition to the spectral properties, Fig.~\ref{SDWfinitesize} provides more information on the finite-size effects of antiferromagnetic spin/charge density wave structure factors (averaged by lattice size). Obviously, significant finite-size effect of $S_{sdw}/N$ indicates the absence of antiferromagnetic magnetic order in the thermodynamic limit while conversely the lack of $S_{cdw}/N$'s size dependence support the existence of the charge density order as expected due to the staggered potential.

\section{Summary}

We have explored the bilayer ionic Hubbard model with two types of ionic potentials focussing on the impact of interlayer hybridization on the phases in ionic Hubbard model in square lattice. It turns out that the interaction-driven Insulator-Metal transition in 2D IHM extends to the bilayer IHM with finite interlayer hybridization. We obtained the finite temperature phase diagram via the spectral properties supplemented with the temperature evolution and finite-size effects of local Green's functions. We argue that only the systems with in-phase ionic potential can host the interaction-driven metallic phase; while the anti-phase ionic potential always induces an insulator, whose electrons with opposite spins can occupy the potential valley alternatively between layers accompanied with charge density order. We further investigated the evolution of the antiferromagnetic structure factors of charge/spin density wave with the hybridization $V$ and their finite-size effects, which indicates that both systems have no long-range antiferromagnetic spin density but charge density order.

We point out that our DQMC simulations cannot conclusively distinguish the ultimate ground state due to the unaffordable larger lattice sizes and lower temperatures so that we denote the unresolved regime of weak $\Delta/t \leq 0.2$ by gray patch in our phase diagram. Nevertheless, we believe that the current finite temperature data still provide stimulating insights on the lower temperature properties. The ultimate ground states for both interaction-driven metallic phase and weak $\Delta$ regime in the thermodynamic limit deserves further study employing other appropriate methods.

The existence of interaction-driven metallic phase due to in-phase ionic potential implies further extension of the ionic Hubbard model from simplest bilayer case to realistic three-dimensional model. With potential connection to two- orbital/band systems with each orbital/band experiencing the same or different staggered potentials, we hope that our investigation of the bilayer IHM provides a steady step towards fully understanding the insulator-metal transition in systems with ionic potentials. Further possible directions include the realization of bilayer ionic Hubbard model with cold fermionic atom gases loaded into an optical lattice, bilayer IHM in other lattice geometries, and the influence of near-neighbor interaction and/or hoppings etc.

\begin{acknowledgments}
We are grateful to R.T. Scalettar for fruitful discussions and acknowledge CSCS, Lugano, Switzerland for computing facilities.
\end{acknowledgments}

\appendix*
\section{Local $G(\tau)$ at potential valley sites}

In Figs.~2-3, maximum entropy method (MaxEnt) is employed for obtaining the spectral function $N_{-}(\omega)$ and furthermore $N_{-}(\omega=0)$ is used for determining the phase boundary (blue dotted curves) in Fig.~1. As a trickly methodology that is sensitive to the quality of original $G(\tau)$ data, MaxEnt introduces the ambiguity into the reliability of the spectral function.

In this appendix we provide original data of local $G(\tau)$ at potential valley sites for the parameter sets in Figs.~2-3 as supplement.

In principle, the spectral gap can be extracted from the large imaginary-time limit of the Matsubara Green's function
\begin{equation}
   \lim\limits_{\tau\rightarrow \infty} G_{\pm}(\tau) = - \lim\limits_{\tau\rightarrow \infty} \sum\limits_{{\bf j}} \langle
c^{\phantom{\dagger}}_{{\bf j}\pm}(\tau) c^{\dagger}_{{\bf j}\pm}(0) \rangle \propto e^{-\Delta \tau}
\end{equation}

At finite temperatures, the slope of $\log G(\tau)$ at $\tau=\beta/2$ provides hints on the spectral gap since larger slope implies larger energy gap. In Fig.~8, the flatness of the curves in panel (a) signal the metallic behavior; while panel (c) implies a transition from an insulator (blue) to a metal (red). Moreover, panels (b) and (d) always indicate the insulating phase. Similarly, in Fig.~9, panel (a) clearly signifies a transition from a metal to an insulator while panel (b) characterizes an insulator. All these behavior of local $G(\tau)$ are consistent with the spectral functions obtained via maximum entropy method in Figs.~2-3.

\begin{figure}
\psfig{figure=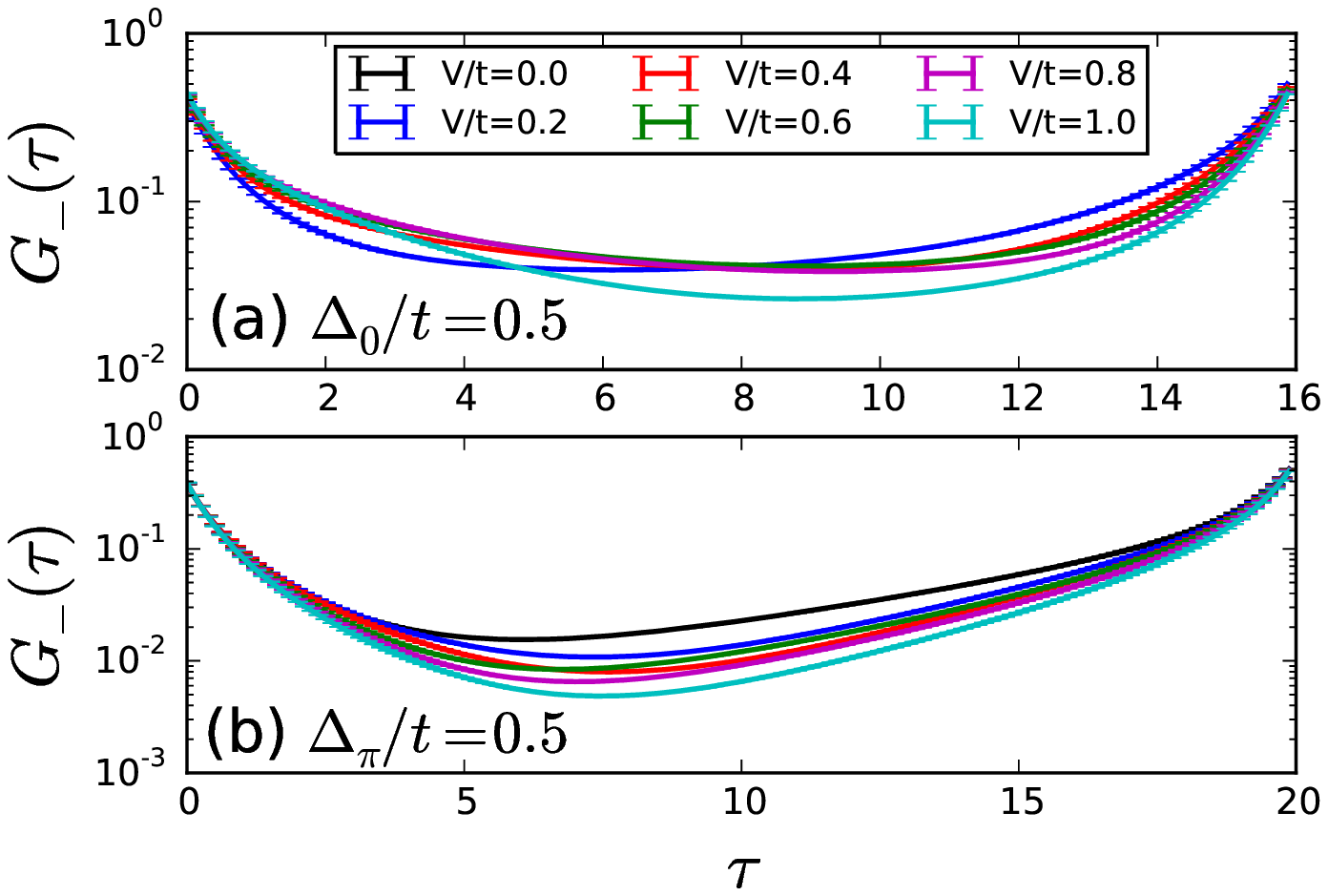,height=5.6cm,width=8.8cm,angle=0,clip,trim={0 0.8cm 0 0}}
\psfig{figure=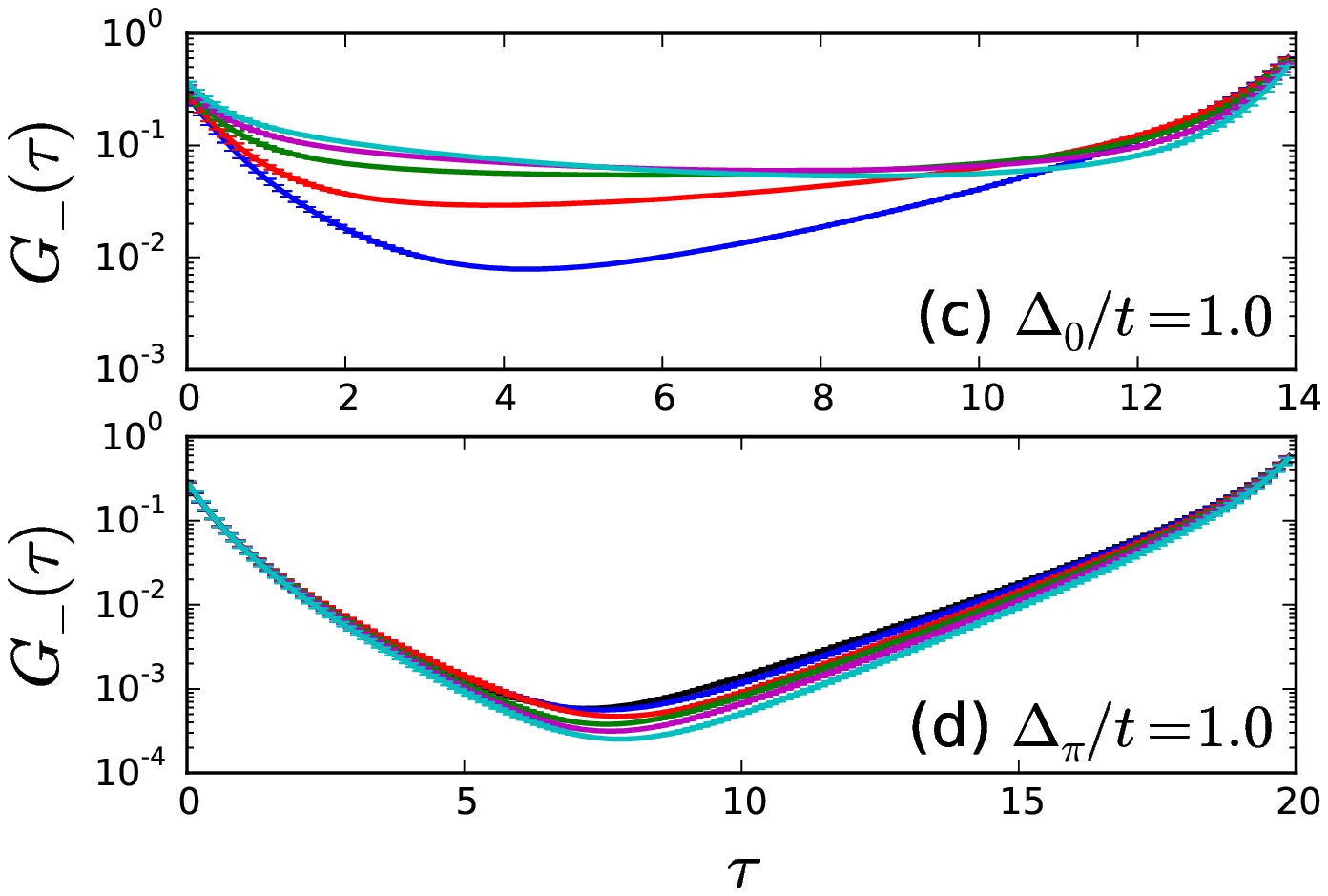,height=5.6cm,width=8.8cm,angle=0,clip,trim={0 0 0 0.7cm}}
\caption{(Color online) Comparison of local $G(\tau)$ at potential valley site between bilayer with in-phase and anti-phase ionic potentials. The curve slope at $\tau=\beta/2$ provides hints on the spectral gap. The flatness of the curves in panel (a) signal the metallic behavior; while panel (c) implies a transition from an insulator (blue) to a metal (red). In addition, panels (b) and (d) indicate the insulating phase.}
\label{Gt1}
\end{figure}

\begin{figure}
\psfig{figure=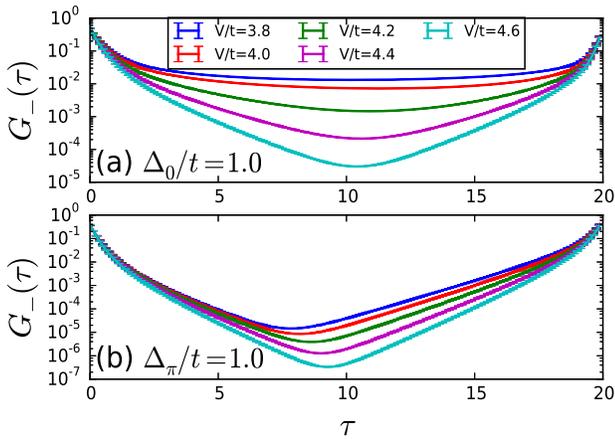,height=6.2cm,width=8.8cm,angle=0,clip}
\caption{(Color online) Similar to Fig.~8 except for the amplitude of ionic potentials. Panel (a) implies a transition from a metal to an insulator while panel (b) characterizes an insulator.}
\label{Gt2}
\end{figure}



\begin{thebibliography}{50}
\bibitem{comment}
The nature of this intermediate phase (bond ordered, metallic etc.) is still in debate. Here we follow the claiming of intermediate metallic phase since the present and all previous DQMC studies generate a finite density of states at Fermi energy, which characters a metallic phase.
\end{thebibliography}
\end{document}